\documentclass{sig-alternate-05-2015}

\usepackage{graphicx}
\usepackage{epsf}
\usepackage{epstopdf}
\usepackage{color}
\usepackage{times}
\newtheorem{thm}{Theorem}
\newtheorem{lem}[thm]{Lemma}

\begin{document}

\CopyrightYear{2016}
\setcopyright{acmcopyright}
\conferenceinfo{MSWiM '16,}{November 13-17, 2016, Malta, Malta}
\isbn{978-1-4503-4502-6/16/11}\acmPrice{\$15.00}
\doi{http://dx.doi.org/10.1145/2988287.2989143}

\title{Holistic Small Cell Traffic Balancing across Licensed and Unlicensed Bands}

\def\sharedaffiliation{
\end{tabular}
\begin{tabular}{c}
}

\numberofauthors{2}
\author{
\alignauthor
Ursula Challita\\
       %\affaddr{School of Informatics}\\
       %\affaddr{The University of Edinburgh}\\
       %\affaddr{10 Crichton St, EH8 9AB, UK}\\
       \email{ursula.challita@ed.ac.uk}
% 2nd. author
\alignauthor
Mahesh K. Marina\\
       %\affaddr{School of Informatics}\\
       %\affaddr{The University of Edinburgh}\\
       %\affaddr{10 Crichton St, EH8 9AB, UK}\\
       \email{mahesh@ed.ac.uk}
       \sharedaffiliation
    \affaddr{School of Informatics}\\
    \affaddr{The University of Edinburgh}}

\maketitle
\begin{abstract}
Due to the dramatic growth in mobile data traffic on one hand and the scarcity of the licensed spectrum on the other hand, mobile operators are considering the use of unlicensed bands (especially those in 5 GHz) as  complementary spectrum for providing higher system capacity and better user experience. This approach is currently being standardized by 3GPP under the name of LTE Licensed-Assisted Access (LTE-LAA). In this paper, we take a holistic approach for LTE-LAA small cell traffic balancing by jointly optimizing the use of the licensed and unlicensed bands. We pose this traffic balancing as an optimization problem that seeks proportional fair coexistence of WiFi, small cell and macro cell users by adapting the transmission probability of the LTE-LAA small cell in the licensed and unlicensed bands. The motivation for this formulation is for the LTE-LAA small cell to switch between or aggregate licensed and unlicensed bands depending on the interference/traffic level and the number of active users in each band. We derive a closed form solution for this optimization problem and additionally propose a transmission mechanism for the operation of the LTE-LAA small cell on both bands. Through numerical and simulation results, we show that our proposed traffic balancing scheme, besides enabling better LTE-WiFi coexistence and efficient utilization of the radio resources relative to the existing traffic balancing scheme, also provides a better tradeoff between maximizing the total network throughput and achieving fairness among all network flows compared to alternative approaches.
\end{abstract}

%\begin{CCSXML}
%<ccs2012>
%  <concept_id>10010520.10010553.10010562</concept_id>
%  <concept_desc>Computer systems organization~Embedded systems</concept_desc>
%  <concept_significance>500</concept_significance>
%  <concept_id>10010520.10010575.10010755</concept_id>
%  <concept_desc>Computer systems organization~Redundancy</concept_desc>
%  <concept_significance>300</concept_significance>
%  <concept_id>10010520.10010553.10010554</concept_id>
%  <concept_desc>Computer systems organization~Robotics</concept_desc>
%  <concept_significance>100</concept_significance>
%  <concept_id>10003033.10003083.10003095</concept_id>
%  <concept_desc>Networks~Network reliability</concept_desc>
%  <concept_significance>100</concept_significance>
%</ccs2012>
%\end{CCSXML}

%\ccsdesc[500]{Computer systems organization~Embedded systems}


%
%  Use this command to print the description
%
%%\printccsdesc

% We no longer use \terms command


\section{Introduction}
The 3rd Generation Partnership Project (3GPP) is considering the deployment of LTE in the 5 GHz unlicensed bands, an approach known as Licensed-Assisted Access using LTE (LTE-LAA)~\cite{edinburgh}, as one of the key mechanisms to cope with the dramatic growth in mobile data traffic as well as the spectrum scarcity problem, especially below 6 GHz. LTE-LAA is an attractive solution for small cells due to the limits on maximum transmit power in unlicensed bands. It will allow the opportunistic use of the unlicensed spectrum as a complement to the licensed spectrum for offloading best-effort traffic via the LTE carrier aggregation (CA) framework, while critical control signalling, mobility, voice and control data will always be transmitted on licensed bands. Therefore, the performance experienced by mobile UEs as well as the utilization of the unlicensed spectrum will be enhanced.

LTE-LAA, however, introduces new and inter-dependent challenges of LTE-WiFi coexistence in unlicensed bands, traffic offloading from licensed to unlicensed spectrum, and inter-operator spectrum sharing in unlicensed bands~\cite{overview}. LTE-WiFi coexistence depends on the extent to which LTE-LAA small cells (operating in both licensed and unlicensed bands) rely on unlicensed spectrum to meet their traffic demand, and this in turn is dependent on the nature of inter-tier interference in the licensed spectrum shared by a macro cell and small cells in its coverage area. This link between LTE small cell operation in the unlicensed band and inter-tier/inter-cell interference in the licensed spectrum is essentially the traffic balancing problem\footnote{Traffic balancing can be seen as addressing LTE-WiFi coexistence and LTE traffic offloading challenges together.} and the focus of this paper. The transmission of the small cell base station (SBS) on the unlicensed band can disrupt WiFi transmissions as the latter relies on a contention-based channel access and hence starvation may occur when co-existing with LTE. On the other hand, LTE-LAA SBS transmission on the licensed band can cause inter-tier/inter-cell interference to the macro cell and other small cell users, potentially degrading their throughput. Thus addressing the traffic balancing problem is challenging as it entails a LTE-LAA small cell base station to adaptively decide on how to steer its traffic between the licensed and unlicensed bands while optimizing the overall network performance and achieving fair coexistence among the technologies operating on both bands. Though the above discussion highlights the importance of traffic balancing for optimizing the performance of co-located networks based on different technologies (LTE and WiFi) sharing same unlicensed bands, and for more effective LTE-WiFi coexistence, this problem has till date received little attention in the research literature with \cite{imp_extended} as the only notable work. Nevertheless, the work in \cite{imp_extended} leads to an inefficient utilization of the available resources due to the inefficient coexistence mechanism on the licensed band as well as the sequential adaptation approach for optimizing both bands, which we further discuss in Section 2.

In this paper, we take a holistic approach for LTE-LAA small cell traffic balancing across licensed and unlicensed bands. In other words, we aim to jointly address the LTE-LAA small cell operation in licensed and unlicensed bands by determining its transmission behavior on both bands in a coordinated fashion depending on the interference/traffic levels on each of the bands. Specifically, we make the following key contributions:

\begin{itemize}
  \item We present a formulation of the optimization problem for holistic traffic balancing that seeks proportional fair coexistence of WiFi, small cell and macro cells by deciding on the transmission probability of LTE-LAA small cell in the licensed and unlicensed bands. The intention behind this formulation is for the LTE-LAA SBS to switch between or aggregate licensed and unlicensed bands depending on the interference/traffic level and number of active UEs in each cell. We derive a closed form solution for the aforementioned optimization problem. An attractive aspect of our solution is that it can be applied online by each LTE-LAA SBS, adapting its transmission behavior in each of the bands, and without explicit communication with WiFi nodes. (Section~\ref{sec:formulation})

  \item We also propose a transmission mechanism for the operation of SBS on the licensed and unlicensed bands. Our mechanism leverages the above mentioned traffic balancing solution and aims at avoiding the disruption to on-going WiFi transmissions while adhering to the LTE frame structure. (Section~\ref{sec:channelaccess})

  \item We provide extensive numerical and simulation results using several scenarios to highlight the main capabilities of our proposed scheme. Results show that LTE-LAA SBS, aided by our scheme, would adaptively steer its traffic from one band to another or transmit on both bands simultaneously depending on the interference/traffic levels and number of active UEs on each of the bands. Simulation results additionally demonstrate the effectiveness of our proposed scheme in comparison with \cite{imp_extended} and other approaches, representing the state-of-the-art. They reveal that approaches focusing on coexistence in one band while ignoring the other cause load imbalance and a decrease in the total network throughput and/or fairness. On the other hand, our approach, aided by its holistic nature, results in improved network performance as it achieves a better tradeoff between maximizing the total network throughput and attaining fairness among all network flows while also providing better LTE-WiFi coexistence. (Section~\ref{sec:evaluation})

      %Note that the resulting gains of our proposed mechanism stem from its holistic nature.
      % whereby SBS jointly optimizes the performance of the UEs on both bands.}
\end{itemize}

%The rest of the paper is organized as follows. The following section discusses related work. Section~\ref{sec:model} details the system model for the coexistence of an LTE-LAA SBS with an LTE MBS and WLANs in the overlapping coverage area. In Section~\ref{sec:formulation}, we present an optimization problem for balancing traffic of the LTE-LAA SBS on the licensed and unlicensed bands and also derive a closed form solution for the problem. Section~\ref{sec:channelaccess} describes our proposed transmission mechanism for the operation of the LTE-LAA SBS on the licensed and unlicensed bands. Section~\ref{sec:evaluation} presents numerical and simulation results for the proposed algorithm and compares its with other approaches. Finally, conclusions are drawn in Section~\ref{sec:conclusion}.
%In Section~\ref{sec:simulation}, we evaluate the proposed algorithm through system simulations.

\section{Related Work}\label{sec:related}
LTE use of unlicensed bands has been receiving growing amount of attention within the research community in recent years. The authors in~\cite{overview} provide an overview of LTE-LAA as well as the benefits and challenges it brings. Several papers have looked at the performance impact of LTE operating in unlicensed bands on WiFi. In a recent paper~\cite{ICC_experimental_evaluation}, the authors conduct an experimental evaluation for characterizing the interference impact of LTE-LAA on WiFi under various network conditions; it is shown that the impact of LTE-LAA on WiFi throughput depends on the channel bandwidth, center frequency and MIMO and can be heavily degraded for some scenarios. Concerning mechanisms for LTE-WiFi coexistence, most of the previous work uses muting (adaptive duty cycling)~\cite{ICC, xing, cano, dyspan, cu_LTE}. More crucially, much of the existing work does not consider the operation of LTE-LAA SBS in the licensed band while optimizing its use in the unlicensed bands alongside WiFi. This can however lead to a suboptimal resource allocation when seen globally. For instance, it can result in an over-utilization of the unlicensed band by LTE-LAA SBS and a decrease in WLAN performance, as it will be shown later in Section~\ref{sec:evaluation}.

LTE-LAA small cells enable efficient and flexible use of the unlicensed spectrum, leveraging the LTE-Advanced carrier aggregation feature. Nevertheless, early work on traffic balancing across licensed and unlicensed bands (e.g., \cite{saad,fujitsu}) focused on dual-access small cells (with both LTE and WiFi air interfaces) and thus lacking these benefits. To the best of our knowledge, \cite{imp_extended} is the only notable traffic balancing work in the literature that applies to LTE-LAA small cells. The proposed traffic balancing technique in \cite{imp_extended} is based on adjusting the power level in the licensed spectrum and the number of muted subframes in the unlicensed bands. We identify three aspects of the work in \cite{imp_extended} discussed below, which together result in a lower WLAN performance and a degradation in the overall network performance compared to our proposed scheme, as shown later in Section~\ref{sec:evaluation}.

\begin{enumerate}
\item {\em Use of power control in the licensed band.} In the context of inter-cell interference coordination (ICIC) management in HetNets, 3GPP Release 10 introduced almost blank subframes (ABS) as an efficient way to enhance the network performance. In~\cite{ABS_1}, the authors evaluate the 3GPP enhanced ICIC (eICIC) techniques through realistic system-level simulations where it is shown that the ABS eICIC time method provides the best macrocell UE (MUE) protection as compared to other eICIC power methods. There is other work (e.g., \cite{muting_power_2}) which also shows that ABS muting achieves better macro-layer performance at less degradation of the SBS layer performance as compared to power adaptation. Therefore, the use of power control on the licensed band in~\cite{imp_extended} leads to a sub-optimal performance on both the licensed and the unlicensed bands given the fact that the coexistence mechanism in the licensed spectrum directly influences the optimization process in the unlicensed band.

\item {\em Considering a fixed level of performance for macrocell base station (MBS).} The use of a fixed and predefined interference threshold value for MBS in~\cite{imp_extended} results in prioritizing the MBS performance irrespective of the degradation level caused to the SBS layer. This uncoordinated optimization approach on the licensed band would result in an unfair share of that band which in turn could lead to an over-utilization of the unlicensed band by the SBS and thus a degradation in the WLAN performance.

\item {\em Sequential approach to optimizing the licensed band first then the unlicensed band.} The authors in~\cite{imp_extended} consider a sequential approach for optimizing both bands i.e., the output of the power allocation sub-problem in the licensed spectrum serves as an input to the muting sub-problem for the unlicensed bands. This results in prioritizing the licensed band and potentially over-utilizing the unlicensed band by SBS as well as degrading the total network performance.
\end{enumerate}

\section{System Model}\label{sec:model}

\begin{figure}[t!]
%\vspace{-0.5in}
  \begin{center}
    \includegraphics[scale=0.21]{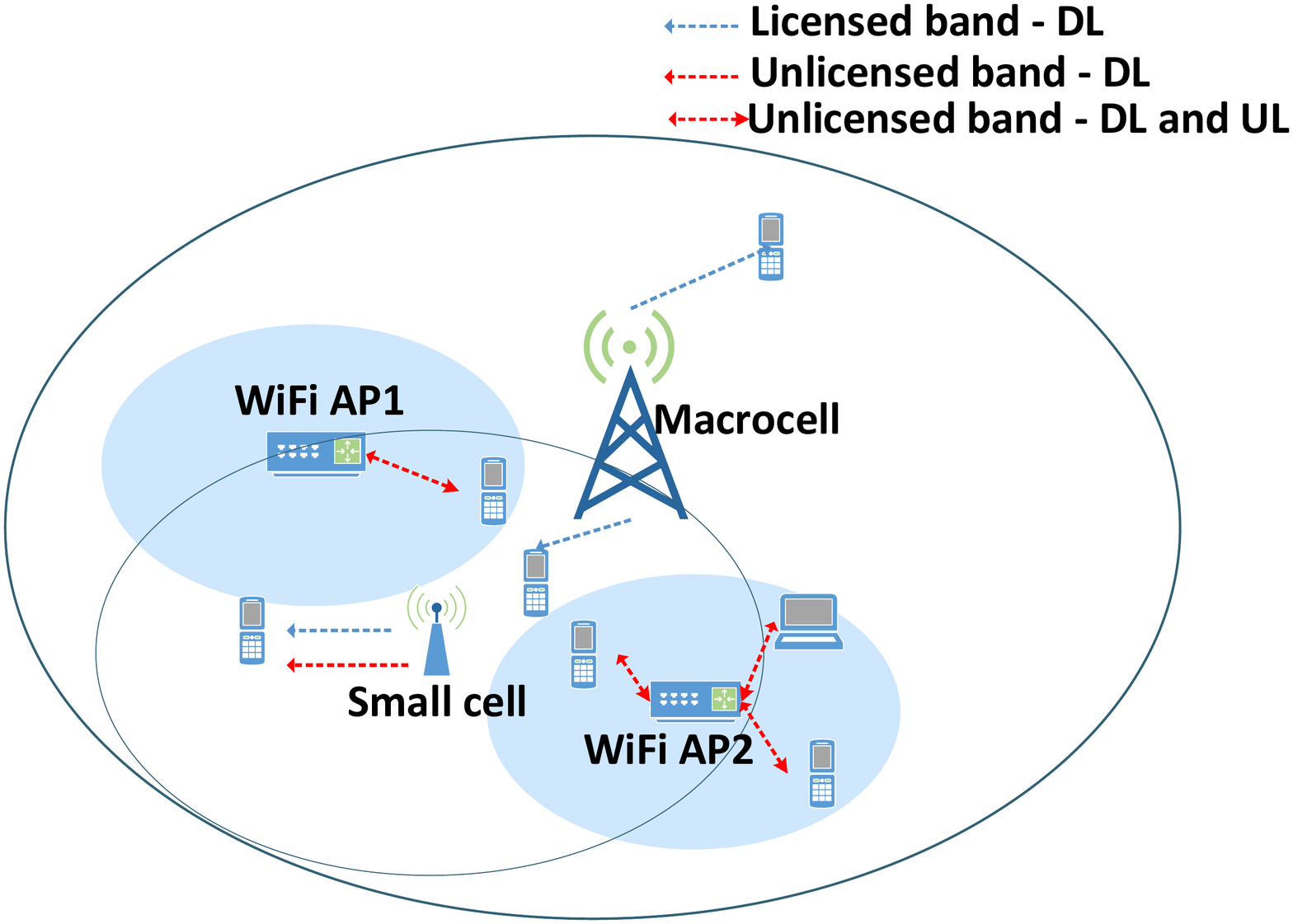}
   \caption{Illustration of the system model.}\label{system_model}
   %\vspace{-0.4cm}
  \end{center}
\end{figure}
%\vspace{-2cm}

We consider a system model (depicted in Figure~\ref{system_model}) similar to that in \cite{imp_extended, cano} consisting of a macrocell base station, a small cell and multiple independently operated WiFi networks. We assume a dual band small cell that transmits on both licensed and unlicensed bands via the LTE carrier aggregation feature. The licensed band is shared between MBS and SBS where smaller portions of the spectrum, referred to as Resource Blocks (RBs), are allocated to UEs. On the other hand, SBS and WiFi networks share an unlicensed channel in the time domain and hence at a particular time, the unlicensed channel is occupied by either SBS or WiFi. This represents a dense WiFi deployment scenario where SBS and WiFi may need to time share the same channel.

Let $N_m$, $N_f$ and $N_w$, respectively, denote the number of macro-cell UEs, small cell UEs (SUEs) and WiFi stations (STAs) in a given time period $T$. We assume the supplemental downlink (SDL) mode for the transmission of the small cell in the unlicensed band. On the other hand, traffic for WiFi STAs can be in either uplink or downlink directions. A full-buffer traffic model is assumed for the SBS, consistent with the motivation for SBS to use both licensed and unlicensed bands to meet its traffic demand.
%On the other hand, we assume a finite-buffer traffic model for the MBS and WiFi APs where the traffic has a continuously changing behavior. %We consider that there are always active WiFi, macro and femto users in the network at different times.

%The main parameters of the considered system model are the following:
%\begin{itemize}
%\item A given set of active WiFi stations (STAs) during a given period $T$, $\mathcal{W}=\{w: w=1,\ldots,N_w\}$.
%\item A given set of active macrocell UEs during a given period $T$, $\mathcal{M}=\{m: m=1,\ldots,N_m\}$.
%\item A given set of active LTE-LAA small cell UEs (SUEs) during a given period $T$, $\mathcal{F}=\{f: f=1,\ldots,N_f\}$.
%\item A given set of WiFi APs, $\mathcal{P}=\{p: p=1,\ldots,P\}$, with cartesian coordinates $(x_p, y_p)$.
%\end{itemize}
%where $N_m$, $N_f$ and $N_w$ are the number of MUEs, SUEs and WiFi STAs respectively.

In order to coexist with MBS on the licensed band and WLAN on the unlicensed band, we adopt in our model a holistic traffic balancing approach where SBS adjusts the proportion of time it transmits on both licensed and unlicensed bands. Therefore, at a particular time, the small cell would adaptively choose to transmit on the licensed, unlicensed or both bands depending on the interference level and traffic load of MUEs and WiFi nodes. The proposed scheme can be implemented at the MAC layer and hence the traffic assignment would be transparent to applications on the UEs. SBS would defer from transmission on the unlicensed band in order to allow WiFi transmission opportunities and on the licensed band in order to avoid inter-tier interference.
%We assume that the femtocell can sense the WiFi traffic and can get information about the interference it causes to nearby macro UEs and hence adjusts its transmission on the licensed and unlicensed bands accordingly. For the licensed band, a femtocell can learn the macrocell throughput and hence the interference it causes to the macro UEs through information exchange over the X2 interface.
Therefore, to decide on the proportion of time the small cell transmits on the licensed and unlicensed bands, the following decision variables are defined:
\begin{itemize}
  \item $\alpha \epsilon [0,1]$: the fraction of time SBS is \emph{muted} on the unlicensed channel.
  \item $\beta \epsilon [0,1]$: the fraction of time SBS is \emph{transmitting} on the licensed band.
\end{itemize}

Note that upon muting on the licensed band, SBS would defer from sending data on the physical channels, however, would still send control and reference signals, an approach known as almost blank subframe~\cite{ABS_1}. On the other hand, the use of unlicensed band by the small cell is limited to data plane traffic while control and reference signals are transmitted by the SBS on a licensed carrier, which is essentially the license assisted access (LAA) aspect of LTE-LAA. Concerning the LTE-WiFi coexistence mechanism in the unlicensed band, even though the work of 3GPP LTE-LAA study group is in the direction of standardizing the listen-before-talk (LBT) mechanism, we choose muting as the coexistence mechanism in this work influenced by two observations: (i) most of the LTE-WiFi coexistence literature focuses on adaptive muting; (ii) recent work in~\cite{cano_comparison} shows that conceptually both LBT and adaptive duty cycling (muting) provide the same level of fairness to WiFi transmissions when properly configured.

%check UE reservation signal transmission on the unlicensed band.

%Alternatively, the SBS can adjust the transmission power based on the CSI feedback of UEs, where the transmission power to edge users can be boosted within the range of regulation requirement, thus the aggressive nodes nearby the victim UE may have no chance to access the channel since it is probable that the sensed power exceed the CCA threshold.}

%The current CSI measurement and report has 4ms CSI feedback delay. In the unlicensed band, the interference fluctuation is more frequent and may result in inaccurate CSI results. E.g., the potential hidden node may be idle when the victim UE performs CSI measurement and may start to occupy the channel after this CSI is reported. Therefore, it should be considered to shorten the CSI feedback delay if CSI measurement and report is adopted to help solve the hidden node problem.

%A possible method is that the interfered UEs can precisely recognize the dominant LAA-LTE interferers (i.e., hidden nodes) and shut them down correspondingly, by either sending a CTS-like signal or ask help from its serving cell to prevent their transmission. How to detect the main LAA-LTE interferers, especially for inter-operator detection needs further consideration.

%Unlicensed band based channel status report along with channel reservation signal can be used for avoiding hidden node problem in view of timeliness and effectiveness of the reported channel information.

\subsection{Throughput Modeling}\label{sec:throughput}
In order to assess the network performance for the coexistence of LTE MBS, LTE-LAA small cell and WiFi, we define the throughput for each of the MUEs, SUEs and WiFi STAs.

Upon the transmission on the licensed band, SBS would share the frequency band with MBS. In LTE, the downlink RB allocation among UEs is via OFDMA, implying no intra-cell interference. However, frequency reuse in LTE can be one where macro and adjacent small cells may transmit on the same frequency leading to inter-cell interference. On the other hand, when SBS is transmitting on the unlicensed channel, it shares the channel with WLAN. Therefore, the downlink SINR at SUE $f$, served by SBS $F$, in our model assuming a single MBS and SBS, during the transmission of SBS on the licensed and unlicensed channels respectively, can be expressed as follows:
\begin{equation}
\Gamma_{F,f}^l=\frac{P_{F,f}}{\sigma^{2}+ I_{M,f}} \qquad\text{and}\qquad \Gamma_{F,f}^u=\frac{P_{F,f}}{\sigma^{2}+ I_{W,f}}
\end{equation}
where $P_{F,f}$ denotes the received signal power for SUE $f$ from its serving SBS $F$, $\sigma^{2}$ is the thermal noise power, $I_{M,f}$ represents the interference power from MBS $M$ on SUE $f$ and $I_{W,f}$ corresponds to the aggregate interference power from neighboring WLAN APs/STAs on SUE $f$. Note that upon the transmission of SBS on the unlicensed channel, WLAN would defer from transmission since WiFi STAs sense the carrier, i.e. listen to the channel before transmissions, and transmit only if the channel is idle. Therefore, $I_{W,f}$ corresponds to the interference power due to WLAN hidden terminals.

Similarly, the downlink SINR at MUE $m$, served by MBS $M$, during the non-ABS and ABS periods of SBS on the licensed band respectively, can be expressed as follows:
\begin{equation}
\Gamma_{M,m}^{\mathrm{noABS}}=\frac{P_{M,m}}{\sigma^{2}+ I_{F,m}} \qquad\text{and}\qquad \Gamma_{M,m}^{\mathrm{ABS}}=\frac{P_{M,m}}{\sigma^{2}}
\end{equation}

where $P_{M,m}$ denotes the received signal power for MUE $m$ from its serving MBS $M$, and $I_{F,m}$ represents the interference power from SBS $F$ on MUE $m$.

We denote by $s_k$ the total throughput attained by an LTE UE $k$ (where $k$ is $m$ or $f$). An upper bound for the downlink UE throughput, based on Shannon's capacity, is computed as follows:
\begin{equation}
s_k(\mathrm{bps})=\mathrm{BW}_k\cdot\mathrm{log}_2(1+\Gamma_{k})
\end{equation}
where $\mathrm{BW}_k$ is the channel bandwidth allocated to UE $k$ and $\Gamma_{k}$ is the SINR value of UE $k$.

To derive the throughput attained by a WiFi STA $w$ when using the unlicensed band exclusively, we consider a slotted channel, as per the IEEE 802.11 modus operandi \cite{802.11:2012}. Let $\tau_w$ denote the stationary probability that station $w$ is attempting transmission in a randomly chosen slot time. The total throughput $\hat{s}_{w}$ attained by a WiFi STA $w$ when using the channel {\em exclusively} is:
\begin{equation}
\hat{s}_{w}(\mathrm{bps})=\frac{P_{w,succ}\cdot E[D_w]}{P_{w,idle} \cdot \sigma + P_{w,busy} \cdot T_b},
\end{equation}
where $E[D_w]$ is the expected payload size for station $w$, $P_{w,succ}$ is the probability of a successful transmission and can be expressed as $P_{w,succ}=\tau_w \prod_{i=1, i\neq w}^{N_w} (1-\tau_i)$, $P_{w,idle}$ is the probability of an idle slot and can be expressed as $P_{w,idle}=\prod_{w=1}^{N_w}(1-\tau_w)$ and $P_{w,busy}$ is the probability of a busy slot, regardless of whether it corresponds to a collision or a successful transmission and can be expressed as $P_{w,busy}=1-\prod_{w=1}^{N_w}(1-\tau_w)$ \cite{Duffy:2010}. $\sigma$ and $T_b$ correspond to the average durations of an idle and a busy slot respectively and thus the denominator corresponds to the mean duration of a WiFi MAC slot.

%To avoid the computation of the stationary distribution of the Markov Chain modelling the 802.11 protocol \cite{Duffy:2010}, we employ a renewal-reward approach and obtain $\tau_w$ as the ratio of expected number of attempts $E(T)$ required to transmit a packet, to the expected number of slots $E(S)$ elapsed until the packet transmission completes. That is
%\begin{equation}
%\tau_w=\frac{E(T)}{E(S)}=\frac{1+p_w+p_w^2+...+p_w^R}{t_w+b_0+p_w b_1+p_w^2 b_2+...+p_w^R b_R},
%\label{eq:tau}
%\end{equation}
%where $R$ is the maximum retry limit, $b_k$ is the mean number of slots at backoff stage $k$, and $t_w$ is the mean idle time that a station waits for a packet
%after transmission. We model the WLAN throughput of a station $w$ with variable load and, neglecting post-backoff and assuming no buffering, we express
%\begin{equation}
% t_w = q_w(1+2(1-q_w)+3(1-q_w)^2+...) = \frac{1}{q_w},
%\end{equation}
%where $q_w$ is the probability that a new frame arrives in a uniform slot time $E_s$. Assuming independent Poisson traffic, we can relate the actual packet arrival rate per second (offered load) $\gamma_w$ to the arrival rate per slot $q_w$ of a given station $w$ by
%\begin{equation}
% q_w = 1 - e^{\gamma_w\frac{E_s}{E[P]}}
%\end{equation}

Therefore, during an epoch $T$, the throughput attained by a macro, small cell and WiFi UE respectively can be expressed as follows:
\begin{equation}
s_m=\beta s_m^{\mathrm{noABS}} + (1-\beta) s_m^{\mathrm{ABS}}
\end{equation}
\begin{equation}
s_f=\beta s_f^\mathrm{l} + (1-\alpha)s_f^\mathrm{u}
\end{equation}
and
\begin{equation}
s_w=\alpha \hat{s}_{w}
\end{equation}
where $s_m$, $s_f$ and $s_w$ are the achieved throughputs of MUEs, SUEs and WiFi STAs respectively during a given period of time $T$. $s_m^{\mathrm{noABS}}$ and $s_m^{\mathrm{ABS}}$ correspond to the throughput achieved by MUE $m$ during the transmission of the SBS on the licensed band and during the ABS period of SBS, respectively. $s_f^\mathrm{l}$ and $s_f^\mathrm{u}$ correspond to the throughput of SUE $f$ during the transmission of SBS on the licensed band and an unlicensed channel, respectively.

\section{Holistic Traffic Balancing}\label{sec:formulation}
In order to maximize the total network throughput while coexisting fairly with other LTE and WiFi cells, we aim in this section at proposing a traffic balancing approach that aims at providing a proportional fair coexistence of WiFi STAs, SUEs and MUEs. The rationale behind this approach is to allow SBS to either switch between or aggregate the unlicensed and licensed bands based on the interference level on each band. This will allow higher throughput for MUEs that are in the vicinity of the SBS when SBS is not transmitting on the licensed band, and similarly, more transmission opportunities for WiFi nodes when SBS is not transmitting on the unlicensed band. Therefore, the utility function can be expressed as the product of the throughputs obtained by SUEs, MUEs and WiFi STAs:
%To obtain fairness among the achieved data rates of the small cell, macro cell and WiFi UEs, we aim at maximizing the product of their data rates and hence the utility function can be expressed as follows:
\begin{equation}
\mathcal{U}= \prod_{m=1}^{N_m} s_m \prod_{f=1}^{N_f} s_f \prod_{w=1}^{N_w} s_w
\end{equation}

$\mathcal{U}$ in turn can be expressed as the summation of the logarithmic function of the achieved rates as given below:
\begin{eqnarray}
\mathcal{U}_{\mathrm{log}}= && \sum_{m=1}^{N_m} \mathrm{log} (s_m) + \sum_{f=1}^{N_f} \mathrm{log} (s_f) + \sum_{w=1}^{N_w} \mathrm{log} (s_w) \nonumber \\
= && \sum_{m=1}^{N_m} \mathrm{log} \Big[ \beta s_m^{\mathrm{noABS}} + (1-\beta) s_m^{\mathrm{ABS}} \Big] \nonumber \\
&&+   \sum_{f=1}^{N_f} \mathrm{log} \Big[ \beta s_f^\mathrm{l} + (1-\alpha)s_f^\mathrm{u} \Big]   \nonumber\\
&&+   \sum_{w=1}^{N_w} \mathrm{log} \Big[ \alpha s_{w} \Big]
\end{eqnarray}

The proposed utility function $\mathcal{U}_{\mathrm{log}}$ corresponds to a proportional fair coexistence of MUEs, SUEs and WiFi STAs. The PF scheduling algorithm has been an attractive allocation criterion in wireless networks since it maintains a balance between maximizing the total network throughput while achieving good fairness among network flows~\cite{PF}. Therefore, our optimization problem is formulated as follows:
\begin{equation}
\max_{\substack{\alpha, \beta}} \; \; \mathcal{U}_{\mathrm{log}} \label{obj}
\end{equation}
subject to \hspace{-0.33cm}
\begin{equation}
\alpha \leq \overline{R}_{w} \hspace{1cm}  \label{cons_1}
\end{equation}
\begin{equation}
\alpha \leq \beta \hspace{1cm} \label{cons_2}
\end{equation}
\begin{equation}
0 \leq \alpha \leq 1, 0 \leq \beta \leq 1   \label{cons_3}
\end{equation}

where $\overline{R}_{w} (\leq 1)$ corresponds to the normalized offered load across all WiFi stations; it can be obtained via long-term channel sensing where SBS would monitor the WLAN activity on the unlicensed band and estimate the average WLAN traffic load. In the above formulation, constraint (\ref{cons_1}) limits the fraction of time SBS is muted on the unlicensed band to the time it is busy due to WiFi activity. In other words, it is to make sure that the unlicensed band is not underutilized. The purpose of constraint (\ref{cons_2}) is to ensure that SBS transmits on either the licensed or the unlicensed channel at any given point in time. Constraints (\ref{cons_3}) limit the range of values variables $\alpha$ and $\beta$ can take.

%it can be estimated based on the offered loads and physical layer data rates of each of the individual stations.

%the average of the WLAN normalized offered load over the $N_w$ links and is defined as $\overline{R}_w=(\sum_{w=1}^{N_w}\overline{\gamma}_w)/N_w$. $\overline{\gamma}_w$ is the normalized offered load per link and is defined as the ratio of the packet arrival rate per second per station $w$, to the channel capacity of link $w$, which is the maximum physical layer data rate that can be used for transmission over that link. Therefore, $\overline{R}_{w}$ is a measure of the network utilization and can be considered as the fraction of time a channel is utilized, i.e., the normalized channel busy time (CBT).
%The objective function (\ref{obj}) corresponds to a proportional fair coexistence of LTE-LAA with LTE and WiFi which jointly optimizes the total network performance over both licensed and unlicensed bands.
%is the WiFi load constraint which guarantees that the proportion of time allocated to the WiFi network is limited by the WLAN arrival rate i.e., the channel usage does not exceed the time determined by the total offered load. This constraint guarantees that the unlicensed spectrum is not underutilized.
%guarantees that $\beta$ is larger than $\alpha$, i.e., the proportion of time SBS is transmitting on the licensed band should be larger than the proportion of time SBS is muted on the unlicensed band. This is to guarantee that, at a given time,

\begin{lem}
$\mathrm{log}(x)$ is concave. It follows that the utility function $\mathcal{U}_{\mathrm{log}}$ is an affine combination of concave functions, and hence is concave. Therefore, the optimization problem defined by~(\ref{obj})-(\ref{cons_3}) is concave since the objective function and the feasible region defined by the constraints are concave and hence a closed form solution can be obtained using the Karush-Kuhn-Tucker (KKT) conditions at optimality~\cite{KKT}.
\end{lem}

%\subsection{Analytical solution}\label{sec:closed_form_solution}
%The proposed optimization problem can be solved using any standard non-linear program (NLP) solver, however, its computational time increases with the network size and hence the optimal solution would not be computed online. Therefore, we solve
Based on the above lemma, we now aim to derive a closed form solution for the optimization problem~(\ref{obj})-(\ref{cons_3}) using the KKT conditions at optimality. The KKT conditions are necessary and sufficient for convex optimization problems and consist of the stationarity, primal and dual feasibility, and complementary slackness conditions~\cite{KKT}. Therefore, the Lagrangian of the optimization problem~(\ref{obj})-(\ref{cons_3}) can be written as follows:
\begin{multline}
\mathcal{L}(\alpha, \beta, \lambda_1, \lambda_2, \lambda_3, \lambda_4, \lambda_5, \lambda_6)=-\mathcal{U}_{\mathrm{total}} + \lambda_1(\alpha - \overline{R}_{w}) \\+ \lambda_2(\alpha - \beta) - \lambda_3 \alpha + \lambda_4(\alpha-1) - \lambda_5 \beta + \lambda_6(\beta-1)
\end{multline}

where $\lambda_1$, $\lambda_2$, $\lambda_3$, $\lambda_4$, $\lambda_5$ and $\lambda_6$ correspond to the lagrangian multipliers of constraints~(\ref{cons_1})-(\ref{cons_3}).

%For instance, for constraints (\ref{cons_1}) and (\ref{cons_3}), we consider the possible combinations when at most one Lagrangian multiplier for a given variable of these constraints is NZ i.e., when at most $\lambda_1$, $\lambda_3$ and $\lambda_4$ is NZ or when at most $\lambda_5$ and $\lambda_6$ is NZ. This is due to the fact that one or the other, or both (or the three of them for $\alpha$) of the multipliers will always be equal to zero since only one of the lower or upper bounds of each of the two variables $\alpha$ and $\beta$ can be active at a given time.

In the {\em first step}, we compute the candidates for an optimal solution pair ($\alpha^*$, $\beta^*$) from the possible combinations of feasible solutions satisfying the stationarity and complementary slackness conditions. Note that the total number of possible combinations for the Lagrangian multipliers is 64 (i.e., $2^6$) where a given multiplier could be either zero (Z) or non-zero (NZ) at an optimal solution. However, for our optimization formulation, only 6 combinations are possible candidates for an optimal solution due to some infeasible and redundant combinations. For instance, the combinations that have $\lambda_4$ and $\lambda_5$ as NZ can be omitted since their corresponding solution is ($\alpha^*$, $\beta^*$) = (1,0), however, this will lead to the violation of constraint (\ref{cons_2}). Similarly, if a constraint has finite values for both lower and upper bounds, one would need to consider the possible combinations when at most one of the Lagrange multipliers for
that constraint is NZ. This is due to the fact that one or the other, or both, of the multipliers will always be equal to zero since only one of the bounds can be active at a time. Therefore, the combinations that have both $\lambda_3$ and $\lambda_4$ or $\lambda_5$ and $\lambda_6$ as NZ can be omitted. Moreover, we impose a non-zero muting period on the unlicensed band (i.e., restrict $\alpha$ to be greater than 0) in order to allow the small cell to sense WiFi activity and number of stations and thus we omit the combinations having $\lambda_3$ as NZ. Based on the above, the 6 candidate solutions for $\alpha^*$, $\beta^*$ and $(\lambda_1^*, \lambda_2^*, \lambda_3^*, \lambda_4^*, \lambda_5^*, \lambda_6^*)$ are as follows:

%As noted in the previous section, we restrict $\alpha$ to be greater than 0 to allow the LTE-LAA small cell to sense WiFi activity.

%as can be observed from the range of values $\alpha$ can take,

%the combinations having $\lambda_3$ as NZ lead to an infeasible solution and can thus be excluded as $\alpha$ cannot be zero; $\alpha$=0 corresponds to either $\overline{R}_w$=0 or $N_w$=0, i.e., no WiFi activity, and hence the problem of inter-technology coexistence would not exist for such scenarios.

\emph{Candidate solution 1: $\lambda$=(NZ,0,0,0,0,NZ)}
\begin{equation}\nonumber
\alpha_1=\overline{R}_{w} \;\;\; \mathrm{and} \;\;\; \beta_1=1
\end{equation}

\begin{equation}\nonumber \label{sol1_lambda1}
\lambda_1=- \sum_{f=1}^{N_f} \frac{s_{f}^{u}}{\beta_1 s_{f}^{l} + (1-\alpha_1) s_{f}^{u}} + \frac{N_w}{\alpha_1}
\end{equation}
\begin{multline}\nonumber \label{sol1_lambda6}
\lambda_6= \sum_{m=1}^{N_m} \frac{(s_{m}^{\mathrm{noABS}}-s_{m}^{\mathrm{ABS}})}{\beta_1 s_{m}^{\mathrm{noABS}} + (1-\beta_1)s_{m}^{\mathrm{ABS}}} \\+ \sum_{f=1}^{N_f}\frac{s_{f}^{l}}{\beta_1 s_{f}^{l} + (1-\alpha_1)s_{f}^{u}}
\end{multline}
%A possible scenario for this solution is when the level of inter-tier interference on the licensed band is low, WLAN load is high and the number of active WLAN STAs is greater than or equal to the number of active SUEs.\\

\emph{Candidate solution 2: $\lambda$=(0,0,0,0,0,NZ)}

$\alpha_2$ corresponds to the solution of the following equation:
\begin{equation}\nonumber
\sum_{f=1}^{N_f} \frac{s_{f}^{u}}{s_{f}^{l} + (1-\alpha_2) s_{f}^{u}} - \frac{N_w}{\alpha_2}=0
\end{equation}

\begin{equation}\nonumber
\beta_2=1
\end{equation}

\begin{multline}\nonumber
\lambda_6= \sum_{m=1}^{N_m} \frac{(s_{m}^{\mathrm{noABS}}-s_{m}^{\mathrm{ABS}})}{\beta_2 s_{m}^{\mathrm{noABS}} + (1-\beta_2)s_{m}^{\mathrm{ABS}}} \\+ \sum_{f=1}^{N_f}\frac{s_{f}^{l}}{\beta_2 s_{f}^{l} + (1-\alpha_2)s_{f}^{u}}
\end{multline}
%A possible scenario for this solution is when the level on inter-tier interference on the licensed band is low and the number of active WLAN STAs is close to the number of active SUEs.\\

\emph{Candidate solution 3: $\lambda$=(NZ,NZ,0,0,0,0)}
\begin{equation}\nonumber
\alpha_3=\overline{R}_{w} \;\;\; \mathrm{and} \;\;\; \beta_3=\overline{R}_{w}
\end{equation}

\begin{multline}\nonumber
\lambda_2= -\sum_{m=1}^{N_m} \frac{(s_{m}^{\mathrm{noABS}}-s_{m}^{\mathrm{ABS}})}{\beta_3 s_{m}^{\mathrm{noABS}} + (1-\beta_3)s_{m}^{\mathrm{ABS}}} \\- \sum_{f=1}^{N_f}\frac{s_{f}^{l}}{\beta_3 s_{f}^{l} + (1-\alpha_3)s_{f}^{u}}
\end{multline}

\begin{equation}\nonumber
\lambda_1=- \sum_{f=1}^{N_f} \frac{s_{f}^{u}}{\beta_3 s_{f}^{l} + (1-\alpha_3) s_{f}^{u}} + \frac{N_w}{\alpha_3}-\lambda_2
\end{equation}
%A possible scenario for this solution is when the level of inter-tier interference on the licensed band is high, WLAN load is medium and the number of active WLAN STAs is larger than the number of active SUEs.\\

\emph{Candidate solution 4: $\lambda$=(NZ,0,0,0,0,0)}
\begin{equation}\nonumber
\alpha_4=\overline{R}_{w}
\end{equation}
$\beta_4$ corresponds to the solution of the following equation:
\begin{equation}\nonumber
-\sum_{m=1}^{N_m} \frac{(s_{m}^{\mathrm{noABS}}-s_{m}^{\mathrm{ABS}})}{\beta_4 s_{m}^{\mathrm{noABS}} + (1-\beta_4)s_{m}^{\mathrm{ABS}}} - \sum_{f=1}^{N_f}\frac{s_{f}^{l}}{\beta_4 s_{f}^{l} + (1-\alpha_4)s_{f}^{u}}=0
\end{equation}

\begin{equation}\nonumber
\lambda_1=-\sum_{f=1}^{N_f} \frac{s_{f}^{u}}{\beta_4 s_{f}^{l} + (1-\alpha_4) s_{f}^{u}} + \frac{N_w}{\alpha_4}
\end{equation}

\emph{Candidate solution 5: $\lambda$=(0,NZ,0,0,0,0)}

$\alpha_5$ is equal to $\beta_5$ and their corresponding value is the solution of the following equation:
{\small\begin{multline}\nonumber
-\sum_{m=1}^{N_m} \frac{(s_{m}^{\mathrm{noABS}}-s_{m}^{\mathrm{ABS}})}{\alpha_5 s_{m}^{\mathrm{noABS}} + (1-\alpha_5)s_{m}^{\mathrm{ABS}}} - \sum_{f=1}^{N_f}\frac{s_{f}^{l}}{\alpha_5 s_{f}^{l} + (1-\alpha_5)s_{f}^{u}} \\+ \sum_{f=1}^{N_f} \frac{s_{f}^{u}}{\alpha_5 s_{f}^{l} + (1-\alpha_5) s_{f}^{u}} - \frac{N_w}{\alpha_5} = 0
\end{multline}}

\begin{equation}\nonumber
\lambda_2=-\sum_{f=1}^{N_f} \frac{s_{f}^{u}}{\beta_5 s_{f}^{l} + (1-\alpha_5) s_{f}^{u}} + \frac{N_w}{\alpha_5}
\end{equation}

\emph{Candidate solution 6: $\lambda$=(0,0,0,0,0,0)}

$\alpha_6$ and $\beta_6$ correspond to the solution of the following two equations:
\begin{equation}\nonumber
\sum_{f=1}^{N_f} \frac{s_{f}^{u}}{\beta_6 s_{f}^{l} + (1-\alpha_6) s_{f}^{u}} - \frac{N_w}{\alpha_6}=0
\end{equation}

\begin{multline}\nonumber
-\sum_{m=1}^{N_m} \frac{(s_{m}^{\mathrm{noABS}}-s_{m}^{\mathrm{ABS}})}{\beta_6 s_{m}^{\mathrm{noABS}} + (1-\beta_6)s_{m}^{\mathrm{ABS}}} - \sum_{f=1}^{N_f}\frac{s_{f}^{l}}{\beta_6 s_{f}^{l} + (1-\alpha_6)s_{f}^{u}}=0
\end{multline}

Note that two more candidate solutions exist for $\lambda$= (NZ,NZ,0,NZ,0,NZ) and $\lambda$= (0,NZ,0,NZ,0,NZ) where $\alpha$ and $\beta$ are both equal to 1. However, we can avoid checking these two candidate solutions as they exist only in the case when $\overline{R}_{w}$=1 and hence their solution matches with that of candidate solution 1. %Note, however, that $\overline{R}_{w}$ can never be greater than 1.

In the {\em second step}, we check the primal and dual feasibility conditions for each of the 6 candidate solution pairs and the pair satisfying these conditions is the optimal solution.

Note that all the candidate solutions are independent of the WiFi throughput $s_{w}$ and hence SBS needs to know only the normalized WiFi offered load as well as the number of active WiFi STAs; the SBS can learn the number of active WiFi STAs based on their corresponding MAC addresses during the sensing period~\cite{imp_extended}. The number of MUEs and their throughput can be conveyed to the SBS through the X2 interface. Using this information, SBS can determine the optimal values for $\alpha$ and $\beta$ {\em locally} when needed.

%\ucc{Note that, in order to solve the coexistence problem on the unlicensed channel, a non-zero WLAN activity is considered and thus the value of $\alpha$ is never zero ($\alpha$=0 corresponds to either $\overline{R}_w$=0 or $N_w$=0, i.e., no WiFi activity). Therefore, a minimum muting period is implicitly imposed on the unlicensed band in our optimization problem which guarantees a minimum WLAN transmission and LTE-LAA sensing period.}
%\ucc{mention that we do averaging over multiple T periods.}
%Also add a note that the traffic load doesn't change abruptly. OR look for the minimum period for sensing WLAN activity and add the constraint that alpha should be larger than $t_{min}$. HOWEVER, need to check the derivation of the closed form solution where some candidates are removed because of alpha=0. In that case just add a note here instead of updating the optimization problem.}

%\subsection{Intuitions Behind the Optimal Solution} 
\section{A Transmission Mechanism for LTE-LAA SBS Operation}\label{sec:channelaccess}
\begin{figure}[t!]
  \begin{center}
    \includegraphics[scale=0.25]{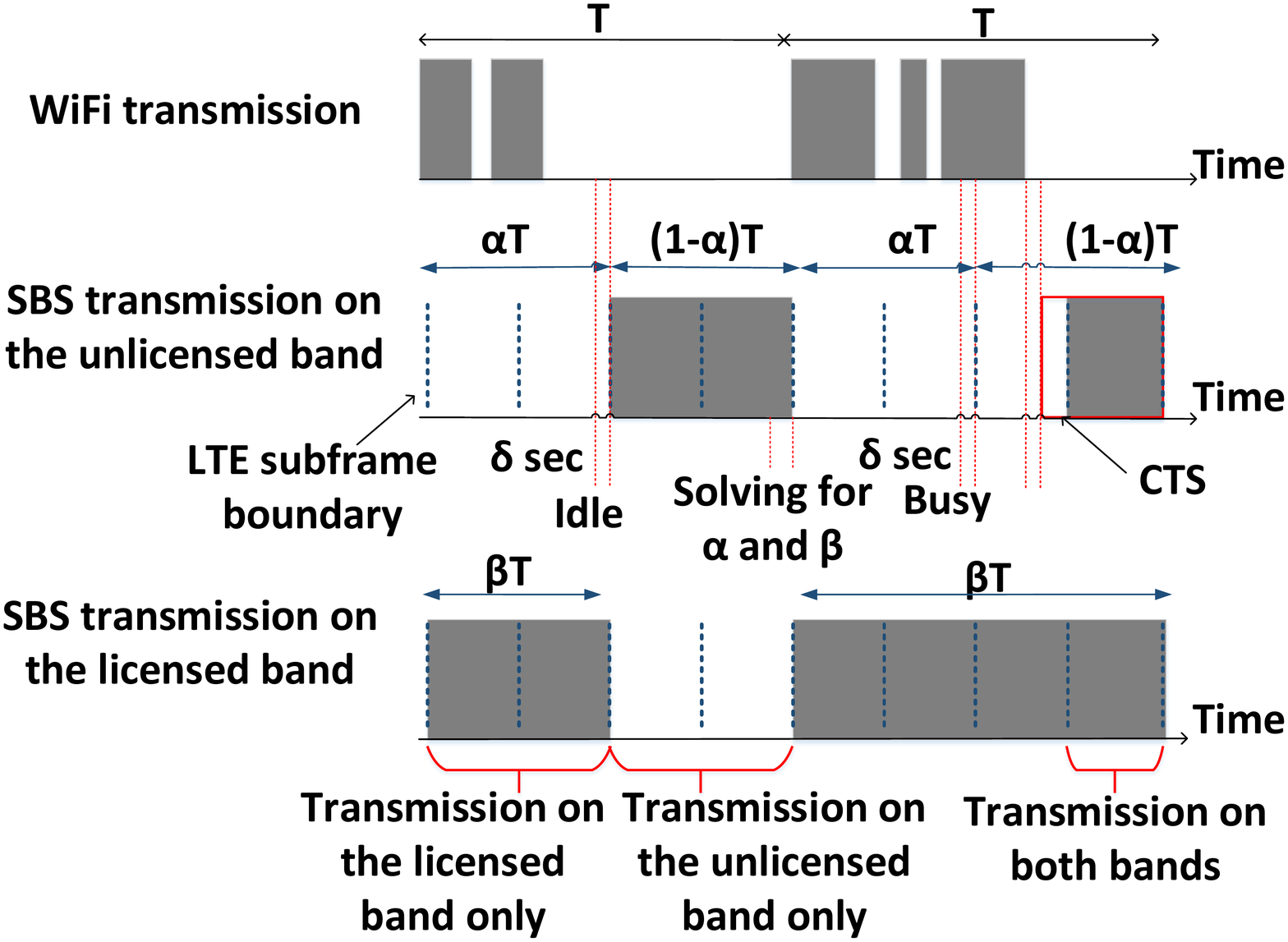}
   \caption{Illustration of the proposed SBS transmission mechanism on the licensed and unlicensed bands. The two possible states upon sensing the unlicensed channel (idle and busy) are demonstrated. SBS will remain in a sensing state when it encounters a busy channel. The three states of SBS (i.e., transmission on the licensed, unlicensed and both bands) are also shown.}\label{channel_access}
   \vspace{-0.35cm}
  \end{center}
\end{figure}

LTE is designed for the exclusive use of the spectrum and hence when operating on the unlicensed band, a new channel access scheme is needed to coexist with other devices having different air interfaces. Therefore, in this section, we propose a transmission mechanism for the operation of an LTE-LAA small cell on the licensed and unlicensed bands. This mechanism builds upon the problem formulation from Section~\ref{sec:formulation} and incorporates a channel access scheme on the unlicensed channel that would allow LTE-LAA SBS to transmit on the unlicensed band in a way that would not disrupt any ongoing WiFi transmissions.

For our proposed mechanism, we divide the time domain into $T$ epochs, where in each epoch we aim at finding the optimal values of $\alpha$ and $\beta$ using the results of Section~\ref{sec:formulation}. Taking into account that LTE transmits only at the beginning of a subframe, our proposed transmission mechanism is aligned with LTE frame structure where $(1-\alpha) T$ and $\beta T$ are rounded to an integer multiple of an LTE subframe duration (1 msec). Moreover, we define $\delta$ as the duration of time the SBS would sense the unlicensed channel before attempting to transmit. Let $\delta$ be such that $\mathrm{SIFS < \delta < DIFS}$, and hence this will guarantee that the ACK of any previous WiFi transmission is received at the sender and that SBS would get access to the unlicensed channel before any other WiFi STA that would be sensing the channel at the same time. The proposed LTE-LAA transmission mechanism is illustrated in Figure~\ref{channel_access} where the two possible states upon sensing the channel (idle and busy) are demonstrated. Moreover, the steps of the proposed mechanism are summarized as follows:

\begin{enumerate}
  \item SBS calculates the values of $\alpha$ and $\beta$ before the beginning of a $T$ period based on the throughput values and number of active nodes of the previous $T$ period and using the results of Section~\ref{sec:formulation}.
  \item At the beginning of a $T$ period, SBS remains silent for the period $\alpha T$ on the unlicensed band and transmits for the period $\beta T$ on the licensed band.
  \item SBS senses the unlicensed channel for $\delta$ sec before $\alpha T$ expires in order to detect any ongoing WiFi transmissions and guarantee alignment with LTE frame structure.
  \item If the channel is idle, SBS transmits for a period of $(1-\alpha)T$.
  \item If the channel is busy, SBS keeps on listening to the channel until it detects a silent period for a duration of $\delta$ sec in order to avoid the disruption to any ongoing WiFi transmission. After detecting a silent period of $\delta$ sec, SBS sends a clear-to-send (CTS) with the duration of the remaining time of the $(1-\alpha)T$ period to reserve the channel for SBS transmission on the unlicensed band. It is important to note that the maximum channel occupancy time is limited to 10 msec after which the unlicensed channel must be released and the LBT process is repeated. Therefore, for the cases where $(1-\alpha)T$ is less than 10 msec, there is a risk that the SBS will not be able to get access to the unlicensed band when the WLAN burst is larger than $(1-\alpha)T$. For such scenarios, the WLAN transmission period for the next $T$ period is shortened accordingly to maintain the average time allocated for LTE-LAA and WLAN.
  \end{enumerate}

\begin{figure}[t!]
  \begin{center}
  %\hspace{-1.5cm}
   \includegraphics[scale=0.21]{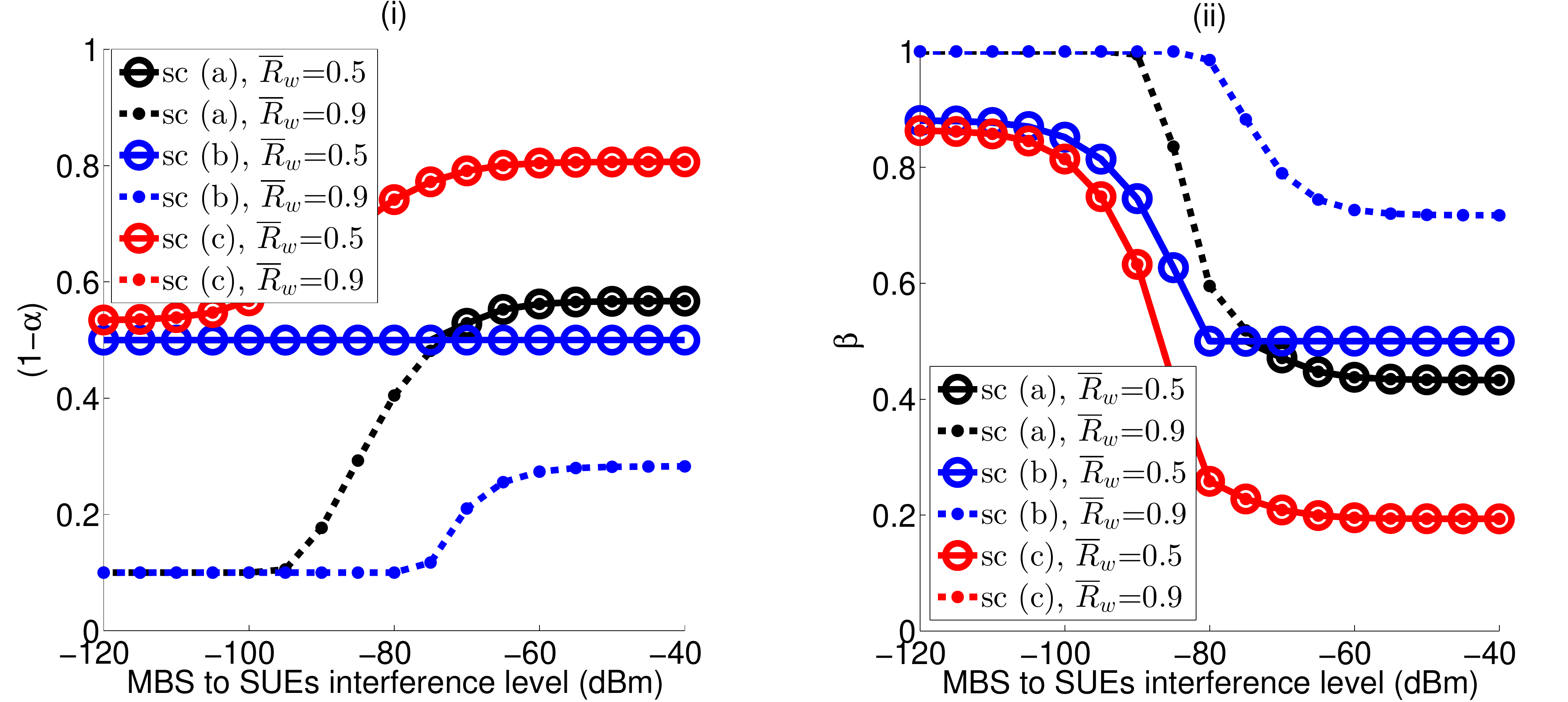}
   \end{center}
   \begin{center}
   \caption{Numerical results for the optimal values of (i) $(1-\alpha)$ and (ii) $\beta$ for varying levels of MBS to SUEs interference in three different scenarios; sc (a) considers an equal number of MUEs, SUEs and WiFi STAs, sc (b) considers the number of WiFi STAs to be three times that of each of MUEs and SUEs and sc (c) considers the number of each of MUEs and SUEs to be three times that of WiFi STAs. For the studied scenarios, we consider medium and high WiFi offered load i.e., $\overline{R}_w$=0.5 and 0.9 respectively, as well as a fixed value for SBS to MUEs interference level (-85 dBm).}\label{numerical}
   \vspace{-0.4cm}
  \end{center}
\end{figure}

\section{Evaluation}\label{sec:evaluation}
%In this section, we aim at providing numerical and simulation results for our proposed traffic balancing algorithm under different network scenarios as well as comparing it to other techniques proposed in the literature.
%In order to validate our numerical results, we aim at conducting simulations for different scenarios and thus extracting insights about our proposed algorithm from both numerical and simulation results.

In this section, we examine the behavior of our proposed holistic traffic balancing scheme in various scenarios using a combination of numerical and simulation results. We also conduct a comparative study of our holistic traffic balancing approach with respect to \cite{imp_extended} and other alternative approaches, representing other proposed techniques from the literature.

In simulations, for WiFi we consider the 802.11 distributed coordination function (DCF) medium access mechanism based on carrier sense multiple access with collision avoidance (CSMA/CA). We assume randomly located STAs that transmit and receive packets according to an independent Poisson process. For simplicity, we consider that all WiFi STAs use the same physical layer parameters, 64-QAM modulation with a 5/6 coding rate when using a 20 MHz channel, which provides a 65 Mbps MAC layer throughput. The simulation parameters for the 802.11 network are the same as those used in~\cite{cano}.

For the LTE and LTE-LAA networks, we assume the same channel conditions for all RBs on both bands and hence the same modulation and coding scheme (MCS) i.e., 64 QAM with 5/6 coding rate, is applied to all RBs of the given 20 MHz channel. Maximum MAC layer throughput for LTE with the above settings is 75 Mbps. These simulation parameters are similar to the ones used in \cite{imp_extended}. We assume a Round Robin (RR) scheduler and equal transmit power for all OFDM symbols in a Transmission Time Interval (TTI) due to the fact that all RBs have the same MCS and thus equal number of bits are allocated to each subcarrier. The maximum transmit power for MBS and SBS is 43 dBm and 23 dBm, respectively. We consider an urban area characterized by the path loss model (for outdoor and indoor locations of the base station and UEs) as given in \cite{pathloss}. A constant payload size of 1500 bytes is assumed for MUEs, SUEs and WiFi STAs. Simulation results are provided for the average of 1000 runs with a 95\% confidence interval.

\subsection{Behavior of $\alpha$ and $\beta$ in different scenarios}
%\subsection{The optimal values of $\alpha$ and $\beta$ as a function of the traffic arrival rate, interference level and number of active UEs}
In this subsection, we study the effect of the variation of the traffic arrival rate as well as the number of active UEs on the values of $\alpha$ and $\beta$ by conducting numerical and simulation results for different practical deployment scenarios.

%\begin{figure*}[t!]
%  \begin{center}
%   \includegraphics[scale=0.45]{figures/numerical_all.eps}%licensed_muting.eps}
%   \caption{The optimal value of $(1-\alpha)$ and $\beta$ versus different values of $s_{f}^{l}/s_{f}^{u}$ and $s_{m}^{\mathrm{noABS}}/s_{m}^{\mathrm{ABS}}$ (a) for a scenario with an equal number of femto, macro and WiFi users (b) for a scenario where the number of WiFi UEs is three times that of femto and macro UEs and (c) for a scenario where the number of macro and femto UEs is three times that of WiFi based on numerical results.}\label{numerical_1}
%  \end{center}
%\end{figure*}

For the numerical results, we consider three different scenarios with different number of MUEs, SUEs and WiFi STAs. Figure~\ref{numerical} shows the optimal values of $(1-\alpha)$ and $\beta$ as a function of the MBS to SUE interference level on the licensed band, for a fixed value of the SBS to MUE interference level (-85 dBm) and two different WLAN traffic loads ($\overline{R}_w$=0.5 and 0.9). Note that the MBS to SUE interference and the SBS to MUE interference levels are relevant during the non-ABS period only.

For the simulation results (shown in Figure~\ref{simulation_1}), we consider only scenario (a) of Figure~\ref{numerical} due to space limitations. Figure~\ref{simulation_1} shows the variation of the proportion of time SBS is transmitting on the licensed and unlicensed bands during the period $T$ as a function of the WLAN traffic arrival rate ($\mathrm{\lambda_{WLAN}}$ (packets/sec)) and for a low and high MUEs traffic arrival rates, i.e., $\mathrm{\lambda_{MUE}}$ = 0.5 and 2 (packets/sec) respectively. Note that $\mathrm{\lambda_{WLAN}}$ and $\mathrm{\lambda_{MUE}}$ correlate to $\overline{R}_w$ and inter-tier interference level respectively of Figure~\ref{numerical}. Each data point in the simulation results is obtained from 1000 runs, each of length 200 msec and with $T$ set to 20 msec.

We can make the following observations from Figures~\ref{numerical} and~\ref{simulation_1}. First, comparing the three considered scenarios of Figure~\ref{numerical}, we conclude that \emph{our proposed traffic balancing scheme provides per node airtime fairness among each of the MUEs, SUEs and WiFi STAs.} For example, consider -60 dBm for the value of MBS to SUEs interference level and $\overline{R}_w$=0.5 for the WLAN load, we observe that in scenario (c), SBS transmits more on the unlicensed band (80\%) and less on the licensed band (20\%) as compared to scenario (b) where SBS transmits 50\% on the unlicensed band and 50\% on the licensed band. This is because the number of each MUEs and SUEs is larger than that of WiFi STAs in scenario (c) while in scenario (b) the number of WiFi STAs is larger than each of the number of MUEs and SUEs.
\begin{figure}[t!]
  \centering
  \includegraphics[scale=0.2]{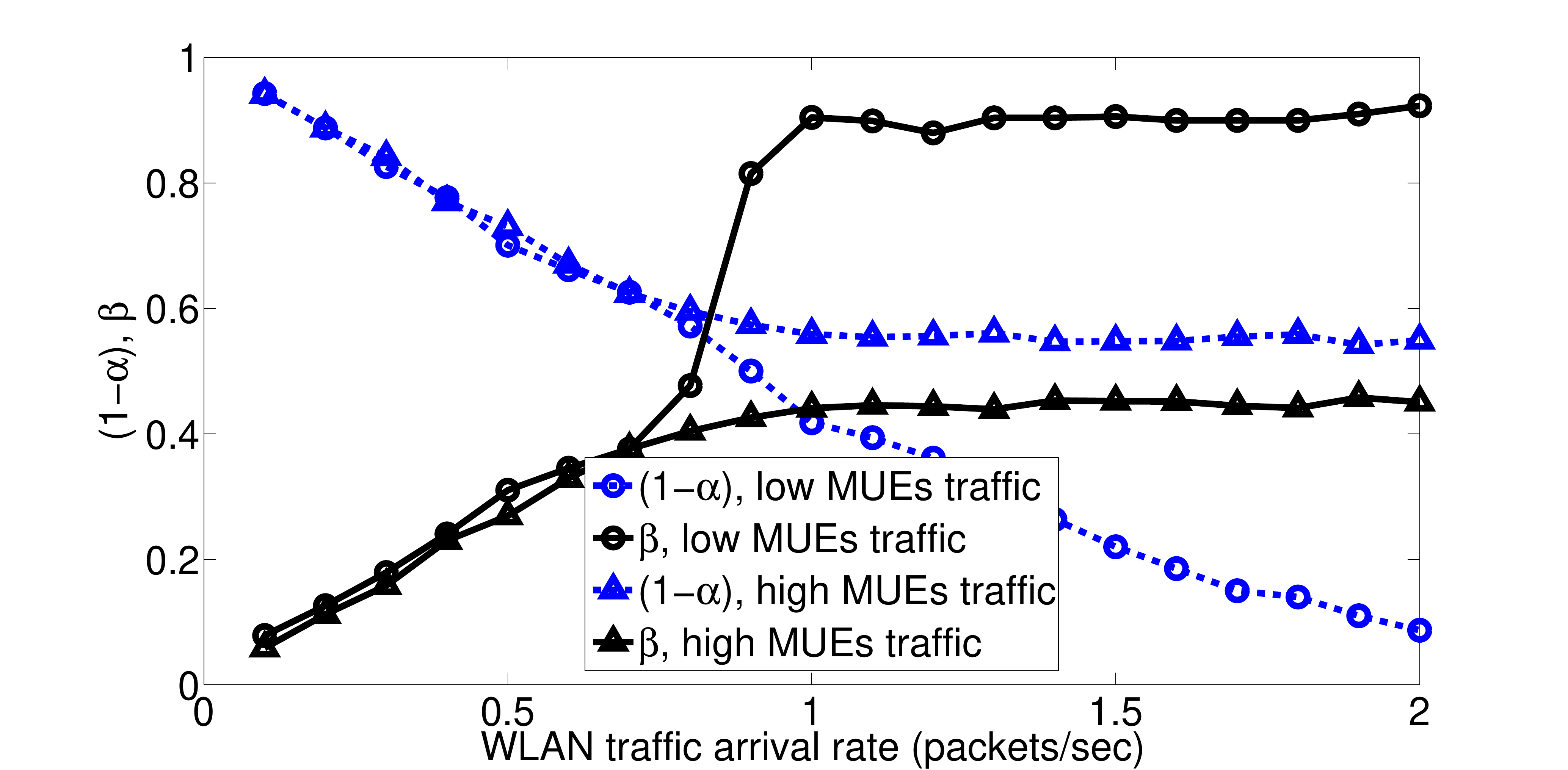}\\
  \caption{Simulation results for the variation of the proportion of time the SBS transmits on the licensed ($\beta$) and unlicensed bands ($1-\alpha$) as a function of the WLAN traffic arrival rate ($\mathrm{\lambda_{WLAN}}$) and for a low and high MUEs traffic arrival rates i.e., $\mathrm{\lambda_{MUE}}$= 0.5 and 2 (packets/sec) respectively, for a scenario of equal number of MUEs, SUEs and WLAN STAs.}\label{simulation_1}
  %\vspace{-0.4cm}
\end{figure}

Second, \emph{our proposed scheme copes with the interference level on both bands by adapting the values of $\alpha$ and $\beta$.} This can be observed for high values of inter-tier interference in Figure~\ref{numerical} or high values of $\mathrm{\lambda_{MUE}}$ in Figure~\ref{simulation_1}. In those scenarios, WLAN shares the unlicensed band with SBS for a proportion of time larger than its idle period, i.e., larger than (1-$\overline{R}_w$), in order to decrease the effect of inter-tier interference on the UEs throughput on the licensed band. For example, in Figure~\ref{numerical}, for scenario (a) and $\overline{R}_w$=0.9, SBS transmits for 55\% of the time on the unlicensed band when the MBS to SUEs interference level is -60 dBm as compared to 10\% when the MBS to SUEs interference level is -95 dBm. This can also be noted from Figure~\ref{simulation_1} where (1-$\alpha$) is equal to 20\% for $\mathrm{\lambda_{MUE}}$= 0.5 (packets/sec) but increases to 55\% for $\mathrm{\lambda_{MUE}}$= 2 (packets/sec), for $\mathrm{\lambda_{WLAN}}$=1.5.

Third, \emph{our proposed traffic balancing scheme allows SBS to transmit on either one of the two bands or aggregate both bands through CA and thus increasing its capacity.} Given that SBS is muted for the period of $\alpha$ and (1-$\beta$) on the unlicensed and licensed bands respectively, we can deduce that it transmits on both bands simultaneously for a period of $(\beta-\alpha) T$ sec, and on one of the two bands for the remaining duration of the $T$ period i.e., for a period of $(1-(\beta-\alpha)) T$ sec, as per our proposed transmission mechanism of Section~\ref{sec:channelaccess}. For example, in Figure~\ref{numerical}, for scenario (b), $\overline{R}_w$=0.5 and MBS to SUEs interference level of -90 dBm, $\alpha$=0.5 and $\beta$=0.75 and thus SBS transmits on both bands simultaneously for 25\% of the $T$ period. This can also be shown in Figure~\ref{simulation_1} where $\alpha$=0.6 and $\beta$=0.9 for $\mathrm{\lambda_{WLAN}}$=1 and $\mathrm{\lambda_{MUE}}$=0.5 and hence SBS transmits on both bands simultaneously for 30\% of the $T$ period.

Fourth, for all the considered scenarios of Figures~\ref{numerical} and~\ref{simulation_1}, we notice that \emph{the unlicensed band is always utilized by either WLAN or SBS and hence this avoids its under-utilization.} In other words, SBS is always transmitting on the unlicensed band for at least the portion of time that it is not utilized by WLAN i.e., (1-$\alpha$) is always greater than or equal to (1-$\overline{R}_w$), consistent with constraint (\ref{cons_1}) in the optimization problem, irrespective of the value of inter-tier interference on the licensed band. For example, for $\overline{R}_w$=0.5 and 0.9, (1-$\alpha$) is always greater than or equal to 0.5 and 0.1 respectively.

Fifth, for all the studied scenarios, \emph{there exists an upper limit for the value of (1-$\alpha$) which corresponds to the maximum proportion of time that WLAN would share its unlicensed band with LTE.} This can be observed in the cases of high inter-cell interference on the licensed band where a minimum airtime portion for WLAN, that is a function of the number of active UEs and WLAN activity, is guaranteed and thus allowing a fair LTE-WiFi coexistence. For example, in Figure~\ref{numerical}, for an equal number of SBS and WLAN UEs (i.e. scenario (a)), the upper limit for (1-$\alpha$) is approximately 0.5.

\emph{Overall, the results demonstrate that our traffic balancing scheme performs as per expectations by steering SBS traffic from one band to another or using both bands simultaneously depending on the level of inter-tier interference on the licensed band, WiFi offered load and number of UEs in each band.}

%As expected, the performance of our traffic balancing algorithm depends on the interference level on both bands and hence the SBS traffic is steered from one band to another based on the level of inter-tier interference and WiFi offered load.

\begin{figure}[t!]
  %\begin{center}
  %\hspace*{-2.5cm}
  \begin{center}
   \includegraphics[scale=0.21]{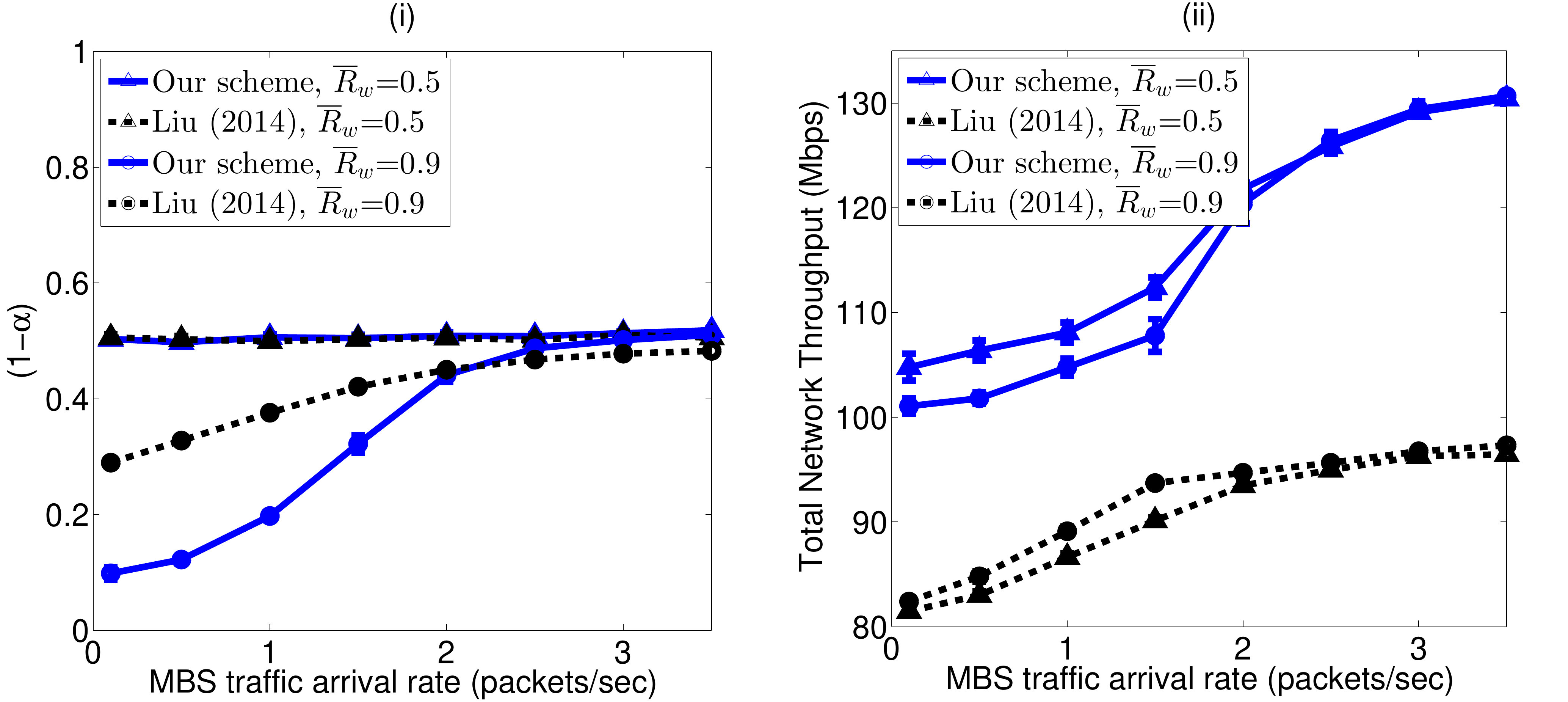} %\hspace*{10cm}
   \caption{Simulation results for (i) the optimal value of the transmission ratio of SBS on the unlicensed band i.e., (1-$\alpha$) and (ii) the total achieved network throughput as a function of the MBS traffic arrival rate ($\lambda_\mathrm{{MUE}}$) for our proposed traffic balancing scheme (Our scheme) and the scheme in \cite{imp_extended} (Liu (2014)). For the comparative study, we consider moderate and high WLAN offered load i.e., $\overline{R}_w$=0.5 and 0.9 respectively.}\label{comparison_LIU}
%   \vspace{-0.35cm}
  \end{center}
\end{figure}

\subsection{Comparison with existing traffic balancing scheme~\cite{imp_extended}}

In this subsection, we compare the performance of our proposed scheme with that of~\cite{imp_extended} which also studies the problem of SBS traffic balancing across licensed and unlicensed bands. Unlike our scheme that jointly optimizes the muting pattern on both bands, the work in~\cite{imp_extended} takes a sequential approach adapting the power level in the licensed band first followed by adjusting the muting pattern on the unlicensed channel. Figure~\ref{comparison_LIU} shows simulation results for (i) the value of (1-$\alpha$) and (ii) the total network throughput for the two schemes as a function of the MBS traffic arrival rate for two different values of the WLAN traffic load ($\overline{R}_w$=0.5 and $\overline{R}_w$=0.9). We can make the following high-level observations from Figure~\ref{comparison_LIU}:
%The results are given for a scenario with an equal number of MUEs, SUEs and WLAN STAs.
%where we divide our analysis into two parts, low and medium MBS traffic arrival rate and high MBS traffic arrival rate.

%\textbf{Observation 1:} \emph{There exists an interdependence between the licensed and the unlicensed bands.}% i.e., the optimization outcome on the unlicensed band is directly influenced by the optimization process on the licensed band.}

\textbf{Observation 1:} \emph{Overall, our proposed traffic balancing scheme achieves better LTE-WiFi coexistence.}

\textbf{Observation 2:} \emph{For all the studied network scenarios, our proposed traffic balancing scheme achieves higher total network throughput.}

In what follows, we examine the reasons behind these observations. First, for scenarios of high WLAN load and when MBS is not in a full buffer state (i.e. $\lambda_\mathrm{{MUE}} <$ 2.5 (packets/sec)), corresponding to candidate solutions 2 or 6, our proposed scheme provides better LTE-WiFi coexistence while also achieving higher total network throughput as compared to~\cite{imp_extended}. This gain is due to the use of subframe muting instead of power adaptation, optimizing the MBS and SBS in a coordinated fashion instead of having a fixed level of performance for MBS, and optimizing the licensed and unlicensed bands in a holistic (joint) manner instead of adopting a sequential approach (see all aspects for \cite{imp_extended} discussed in Section~\ref{sec:related}). The gain for solving the problem holistically as compared to sequentially is characterized separately in Section~\ref{sec:evaluation_comparison} where we consider a variant of our scheme that adopts an independent muting strategy on both bands. On the other hand, the gain due to the other two differences between our scheme and that of~\cite{imp_extended} can be clearly seen from the value of $\alpha$ for candidate solutions 2 or 6 with $N_f$=1:% (which also corresponds to the analytical solution of the proposed formulation in~\cite{imp_extended} ):

\begin{equation}\label{alpha_comparison}
\alpha=\frac{N_w(T_f^l + s_f^u)}{s_f^u(N_w+1)}
\end{equation}

where $T_f^l$ is the throughput achieved by SBS on the licensed band and corresponds to $\beta \cdot s_f^L$ for our proposed scheme and $s_f^L(P_f^*)$ (i.e., a function of the optimal allocated power) for the proposed algorithm of~\cite{imp_extended}. Therefore, from Equation (\ref{alpha_comparison}), we can note that higher values of $T_f^l$ result in higher values for $\alpha$ and thus less utilization of the unlicensed band. Given that ABS muting achieves better macro-layer performance at less degradation of the SBS layer performance as compared to power adaptation, for a specified level of performance for MUEs (e.g., minimum outage level, minimum interference level from SBSs to MUEs), ABS muting causes less degradation in the performance of the SBS layer as compared to power control, i.e., $\beta \cdot s_f^L > s_f^L(P_f^*)$. Following Equation (\ref{alpha_comparison}), our proposed scheme results in less utilization of the unlicensed band and thus allows more WLAN transmission opportunities as compared to~\cite{imp_extended} while maximizing the total network performance.

On the other hand, in the case of a full-buffer MBS (i.e. $\lambda_\mathrm{{MUE}} \geq$ 2.5 (packets/sec)) and at high WLAN load, corresponding to candidate solution 5, we can notice that the value of (1-$\alpha$) for our proposed scheme (0.51) is slightly higher than that of~\cite{imp_extended} (0.49). This is due to the high interference level on the licensed band and thus the need to steer more traffic on the unlicensed band in order to guarantee that the SBS is transmitting on at least one of the two bands at a given time (see constraint (\ref{cons_2}) in the optimization problem). Note, however, that (1-$\alpha$) would converge to its upper limit (i.e., $\sim$ 0.5 for the studied scenarios) and thus allowing a fair LTE-WiFi coexistence.

Second, our proposed scheme achieves similar performance on the \emph{unlicensed} band as that of~\cite{imp_extended} for the case of moderate WLAN load ($\overline{R}_w=0.5$) but it results in a higher total network throughput. For these scenarios, the value of $\alpha$ is limited by $\overline{R}_w$ (corresponding to candidate solutions 1, 3 or 4) and thus the increase in the total network throughput is due to the improvement in the performance on the licensed band i.e., due to the use of subframe muting instead of power adaptation and optimizing the MBS and SBS in a coordinated fashion instead of having a fixed level of performance for MBS (i.e., see aspects (1) and (2) of \cite{imp_extended} discussed in Section~\ref{sec:related}).

In summary, our proposed scheme achieves  better utilization of the available resources compared to \cite{imp_extended} (an increase of 28.3\% in the total network throughput for the studied scenarios) while increasing the transmission opportunities for WiFi on the unlicensed band.

%Therefore, we can conclude that, as compared to~\cite{imp_extended}, our proposed traffic balancing algorithm optimizes the licensed and the unlicensed band in a more efficient way i.e., it achieves better total network throughput while increasing the WLAN transmission opportunities on the unlicensed band.

\subsection{Comparison with alternative approaches}\label{sec:evaluation_comparison}
In this subsection, we compare the performance of our proposed traffic balancing approach with a broad spectrum of alternative approaches. As performance metrics, we consider throughput and fairness obtained using each of the various different approaches. Denote by $\eta(s_i)$ the efficiency of a resource allocation scheme where $\eta(s_i)$ is defined as the sum of all the UEs throughput i.e., $\eta(s_i)$= $\sum_{i=1}^Ns_i$ (where $i$ is $m$, $f$, or $w$ and $N$=$N_m$+$N_f$+$N_w$), and its fairness is given by the Jain's index defined below~\cite{jain}:
\begin{equation}
\mathcal{J}(s_i)=\frac{\big(\sum_{i=1}^N s_i\big)^2}{N\cdot\sum_{i=1}^N s_i^2}
\end{equation}

The value of the Jain's fairness index lies in [$\frac{1}{N}$, 1] where the value of ($\frac{1}{N}$) corresponds to the least fair allocation in which only one UE attains a non-zero throughput and the value of (1) corresponds to the most fair allocation in which all UEs achieve equal rates. Therefore, an efficient allocation of the radio resources seeks to provide a tradeoff between $\eta(s_i)$ and $\mathcal{J}(s_i)$~\cite{jain}.

We compare the throughput and fairness of our proposed scheme with the following set of approaches:
\begin{itemize}
  \item \textit{Case 1 - No Muting on Licensed}: SBS operates on both bands, however, considering a PF muting strategy on the unlicensed band only and hence providing a coexistence technique with WLAN only. On the licensed band, MBS and SBS transmit simultaneously, %i.e., $\beta$=1,
      and hence inter-tier interference is not eliminated.
  \item \textit{Case 2 - No Muting on Unlicensed}: SBS operates on both bands, however, considering a PF muting strategy on the licensed band only and hence providing a coexistence technique with MBS only. On the unlicensed band, SBS is transmitting all the time, %i.e., $\alpha$=0
      and hence excluding any opportunity for WiFi transmissions.
  \item \textit{Case 3 - No Transmission on Licensed}: SBS operates on the unlicensed band only and shares the spectrum with WLAN by muting adaptively. This corresponds to previously suggested approaches such as the work proposed in~\cite{ICC, xing, cano, dyspan, cu_LTE}. For this case, we specifically consider a muting pattern based on PF coexistence of SBS and WLAN on the unlicensed band which is similar to \cite{cano}.
      %This case can be realized through our formulation by setting $\beta$ to 0 and considering a muting based PF coexistence of SBS on the unlicensed band.
  \item \textit{Case 4 - No Transmission on Unlicensed}: SBS operates on the licensed band only and shares the spectrum with MBS by muting adaptively. This corresponds to previously suggested approaches in the area of ICIC such as the work proposed in~\cite{ABS_1} based on muting (ABS). For this case, we specifically consider a muting pattern based on PF coexistence of MBS and SBS on the licensed band.
      %\ucc{For the sake of comparison with our proposed algorithm, we consider a muting pattern based on PF coexistence of SBS and MBS on the licensed band.}
      %We can realize this case by setting $\alpha$ to 1 and considering a muting based PF coexistence of SBS on the licensed band.
  \item \textit{Case 5 - Independent Muting}: SBS operates on both bands, however, an independent mechanism is applied on each band for its coexistence with LTE and WLAN i.e., the coexistence of SBS and MBS on the licensed band and the coexistence of SBS and WLAN on the unlicensed band are solved separately. To realize this case, we consider two independent PF coexistence formulations for the muting of SBS on each of the licensed and unlicensed bands. In other words, when solving for $\alpha$, we consider the WLAN and SBS throughput on the unlicensed band only, and when solving for $\beta$, we consider the MBS and SBS throughput on the licensed band only.
\end{itemize}

\begin{figure}[t!]
  %\begin{center}
  %\hspace*{-3.3cm}
  \begin{center}
   \includegraphics[scale=0.2]{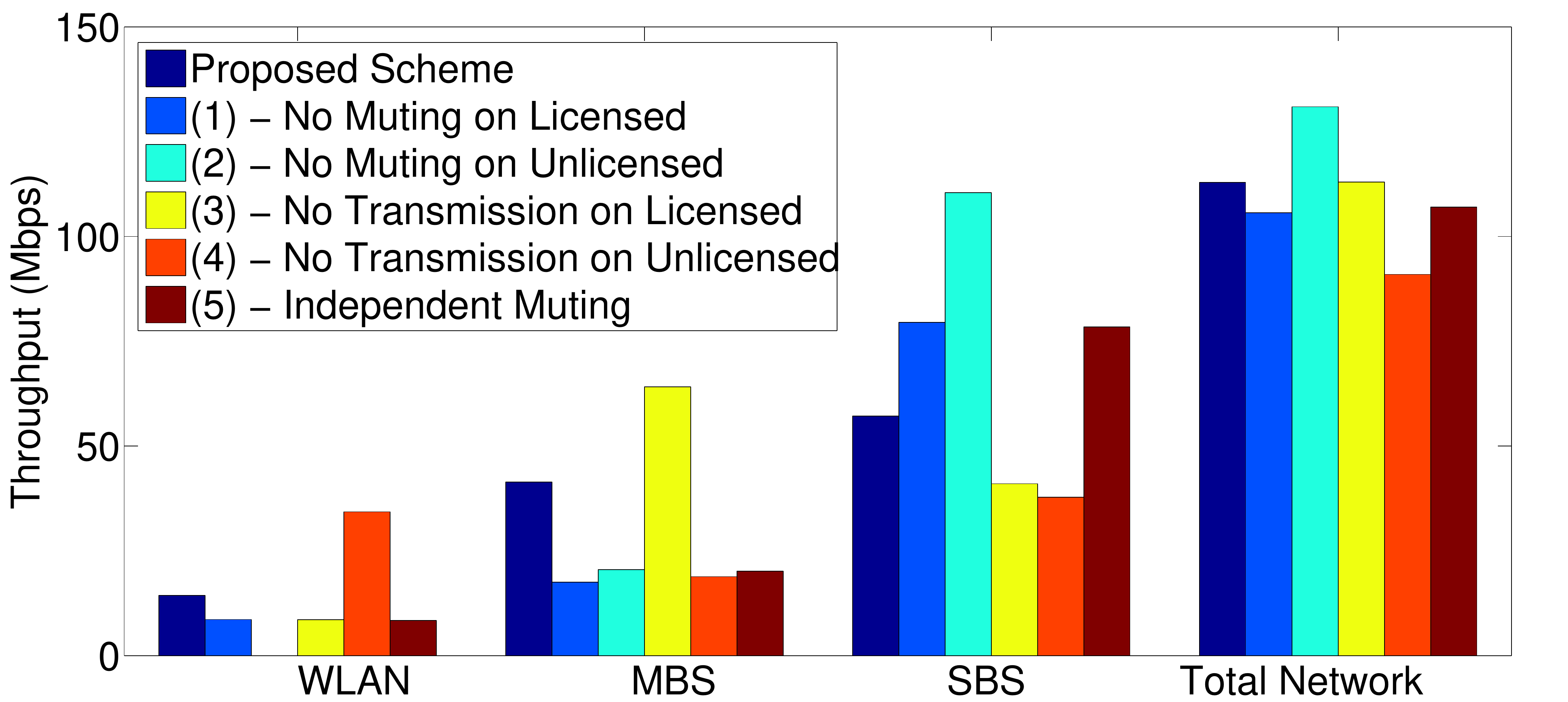} %\hspace*{10cm}
   \caption{The aggregate throughput of the WLAN, MBS, SBS and total network for our proposed traffic balancing scheme in comparison with other approaches.}\label{simulation_comparison}
   \vspace{-0.3cm}
  \end{center}
\end{figure}

Note that cases 1 and 2, respectively, do not consider coexistence mechanisms on the licensed and unlicensed bands and thus are not practical solutions; however, we include them in our study for the sake of completeness.

%Results are provided for the average of four scenarios; scenario 1: medium MBS and WLAN traffic, scenario 2: low MBS and high WLAN traffic, scenario 3: high MBS and low WLAN traffic, scenario 4: high MBS and WLAN traffic.

Figure~\ref{simulation_comparison} shows the throughput achieved by WLAN, MBS, SBS and the total network for our proposed scheme as well as the other five studied approaches; the corresponding Jain's fairness index $\mathcal{J}(s_i)$ values are given in Table~\ref{jain_index}. We can make the following observations from these results. First, the WLAN throughput can be improved when coexisting with LTE-LAA small cells on the unlicensed band by taking into account the transmission of LTE-LAA small cells on the licensed and unlicensed bands and considering a holistic approach for the allocation of the resources on both bands i.e., optimizing both bands jointly. This can be observed from Figure~\ref{simulation_comparison} by comparing the total achieved throughput of WLAN for our proposed scheme with that of cases 1, 2, 3 and 5. Similarly, MBS throughput is higher with our proposed scheme compared to cases 1, 2, 4 and 5. Note that the WLAN and MBS throughputs will be, respectively, maximum when they exclusively use the unlicensed (case 4) and licensed bands (case 3), due to the absence of inter-technology interference in the former and lack of inter-tier interference in the latter. However, the total network throughput is the lowest for case 4; and case 3 results in a relatively unfair sharing of the radio resources as compared to our proposed scheme.

Second, considering an independent muting mechanism on the licensed and unlicensed bands (case 5) leads to performance degradation in terms of throughput and fairness, indicating that the effectiveness of our proposed traffic balancing scheme stems from its holistic nature. This is validated from Figure~\ref{simulation_comparison} and Table~\ref{jain_index} by comparing the total network throughput and Jain's fairness index of our approach to that of case 5 i.e., $\mathcal{J}(s_i)$=0.82 and 0.57 respectively and 5.5\% improvement in the total network throughput. As another observation, the independent muting approach provides very close performance for MBS to case 4 due to the fact that $\alpha$=1 and hence the optimization problem would be a function of the variable $\beta$ only and would correspond to the sub-problem of the coexistence on the licensed band of case 5. Similar argument applies for the WLAN throughput of case 5 which is similar to that of case 3 (where $\beta$=0).

Third, our proposed traffic balancing scheme utilizes the radio resources in the most efficient way compared to the other studied schemes as it provides a better tradeoff between efficiency (throughput) $\eta(s_i)$ and fairness $\mathcal{J}(s_i)$. In terms of efficiency, case 2 provides the maximum total network throughput since SBS will be transmitting on both bands simultaneously, however, WLAN would not be given opportunities for transmission and hence this would result in the least value of $\mathcal{J}(s_i)$ (0.45) as the radio resources are not shared fairly among the different technologies. Note also that our proposed scheme provides similar throughput as case 3; the major contribution to overall throughput in case 3 comes from MBS throughput which is maximum due to its exclusive use of the licensed band. However, comparing Jain's index fairness of our approach to that of case 3, we observe that our scheme allocates the radio resources in a more fair way unlike case 3 that causes a degradation in the WLAN and SBS throughputs. In terms of fairness, case 4 provides the most fair allocation of the licensed and the unlicensed bands as $\mathcal{J}(s_i)$ is the closest to 1 but it comes at the expense of throughput efficiency; total network throughput is the lowest with case 4. The reason for this high value of $\mathcal{J}(s_i)$ is because WLAN would have more transmission opportunities and hence its throughput would increase when using the channel exclusively as compared to sharing it with LTE-LAA SBS. On the other hand, the decrease in the value of $\eta(s_i)$ is due to the difference in the MAC layer throughputs with WiFi and LTE (65 Mbps and 75 Mbps respectively in our simulation setup) and the inter-tier interference level on the licensed band which results in the degradation of the SBS and MBS throughput.
\begin{table}[t!]
   \renewcommand{\arraystretch}{1}
   \caption{Jain's fairness index for the UEs achieved throughput of our proposed scheme and the other five cases.}\label{jain_index}
   \begin{center}
   %\vspace{-0.5cm}
\begin{tabular}{|c|c|c|c|c|c|c|}
\hline
\textbf{Cases} & \textbf{Our scheme} & \textbf{(1)} & \textbf{(2)} & \textbf{(3)} & \textbf{(4)} & \textbf{(5)}\\
\hline
\textbf{$\mathcal{J}(s_i)$} & 0.82 & 0.55 & 0.45 & 0.73 & 0.92 & 0.57 \\
\hline
\end{tabular}
%\vspace{-0.35cm}
\end{center}
\end{table}

%{\bf TODO: need a better more convincing explanation here that matches with your simulation setup. The reason for this high value of $\mathcal{J}(s_i)$ and low value of $\eta(s_i)$ is due to the lower peak data rate of WiFi as compared to that of LTE, for example, TDD-LTE capacity is 1.09 Gbps while that of WiFi is 0.75 Gbps for 4x4 MIMO and 80 MHz channel. Therefore, using the unlicensed band exclusively allows WLAN to acquire more BW, increase its capacity and hence achieve better throughput than the case when it is sharing the spectrum with LTE-LAA, however, resulting in a lower total network throughput.}

%This in turn will result in a higher fairness index among the different network flows, however, less total network throughput.

%this will lead to similar data rates for LTE UEs and WiFi STAs and thus maximizing fairness among different network flows.

%efficiency of the performance of WiFi as compared to that of LTE (e.g., LTE supports higher MCSs and achieves higher peak data rates for the same MCS, WiFi relies on a contention-based channel access protocol and hence performs poorly in the presence of high number of STAs)

\section{Discussion}

In this section, we briefly discuss a couple of issues that warrant detailed exploration in future work.

\subsection{Multiple Channels}

Although we focus on a single unlicensed channel, our traffic balancing scheme can be extended to multiple unlicensed channels, each with a different muting variable \{$\alpha_1$, ..., $\alpha_c$\}, provided that the WiFi networks occupy disjoint channels (non-overlapping channels). Note that in such scenarios, the computational complexity increases due to the increase in the number of variables and thus would make it hard to obtain an online solution. An efficient extension to multiple channels is a key aspect for future work where one could potentially combine channel selection (as studied in \cite{channel_selection_1, channel_selection_2}) with the work in this paper in a joint framework.

\subsection{Hidden Terminals}
%Hidden and exposed terminals are a major problem in wireless networks and can result in a dramatic throughput degradation, if not managed.
LTE use of unlicensed bands in the SDL mode gives rise to hidden terminal situations that need to be handled. In WLAN, this issue is addressed via the request-to-send/clear-to-send (RTS/CTS) messages; however, this method cannot be used for LTE-LAA since only DL transmissions are supported and hence SUEs are not able to transmit the CTS on the unlicensed spectrum. Therefore, to solve the hidden node problem, device-assisted enhancements need to be considered along with other existing mechanisms of the LTE system such as the periodic transmission of UE CSI/interference measurement over the licensed band. On the unlicensed band, a hidden terminal can be detected if SBS senses a good channel while the CSI report from the SUE shows a high interference value. This allows SBS to perform scheduling changes prior and during its operation on the unlicensed channel i.e., exclude the victim SUE for scheduling until its channel becomes idle and schedule other SUEs meanwhile. Alternatively, SBS may select another unlicensed channel to operate on~\cite{hidden_node}.

%Moreover, an SBS can detect whether it experiences exposed terminal problem in the case when it senses a bad channel quality while the CSI from the UE report detects a low interference level.

%Note, that RTS/CTS messages can be exchanged over the licensed band, however, this would incur an additional delay due to the timing constraint between the eNB scheduling and the actual UL transmission from UE on the licensed band (e.g. 4 ms).

\section{Conclusion}\label{sec:conclusion}
In this paper, we have presented a formulation of the holistic LTE-LAA SBS traffic balancing across the licensed and unlicensed bands as an optimization problem that seeks to achieve a proportional fair coexistence of WiFi STAs, SUEs and MUEs. We have derived a closed form solution for the aforementioned optimization problem and proposed a transmission mechanism for the operation of the LTE-LAA SBS on both bands. Results show that LTE-LAA SBS aided by our solution would switch between or aggregate the licensed and unlicensed bands based on the interference/traffic level and number of active UEs in each band. It also provides a better performance for WLAN when coexisting with LTE and an efficient utilization of the radio resources compared to alternative approaches from the literature as it allows a better tradeoff between maximizing the total network throughput and achieving fairness among all network flows.

\section{Acknowledgement}
We would like to thank Paul Patras for his discussions during the early stages of this research work.
%The incentive for the proposed formulation is for LTE-LAA SBS to switch between or aggregate licensed and unlicensed bands based on the interference/traffic level and number of active UEs in each band.

%Future research directions include the extension of our system model to multiple small cells of different operators as well as the transmission of LTE-LAA cells across different channels on the unlicensed band.

\bibliographystyle{abbrv}
\bibliography{references}

\end{document}